\documentclass[twocolumn,showpacs,aps]{revtex4}

\newcommand{\ltsim}{\mbox{{\raisebox{-0.4ex}{$\stackrel{<}{{\scriptstyle\sim}}
$}}}}

\usepackage{graphicx}
\usepackage{dcolumn}
\usepackage{amsmath}
\begin{document}

\title{ Cyclotron resonance harmonics in the organic superconductor
$\beta^{\prime\prime}$-(BEDT-TTF)$_2$SF$_5$CH$_2$CF$_2$SO$_3$;
observation of a new kind of effective mass renormalization }

\author{R.S.~Edwards}
\affiliation{Department of Physics, University of Oxford,
The Clarendon Laboratory,
Parks Road, Oxford,~OX1~3PU, U.K.}

\author{J.A.~Symington}
\affiliation{Department of Physics, University of Oxford,
The Clarendon Laboratory,
Parks Road, Oxford,~OX1~3PU, U.K.}

\author{A.~Ardavan}
\affiliation{Department of Physics, University of Oxford,
The Clarendon Laboratory,
Parks Road, Oxford,~OX1~3PU, U.K.}

\author{J.~Singleton}
\affiliation{Department of Physics, University of Oxford,
The Clarendon Laboratory,
Parks Road, Oxford,~OX1~3PU, U.K.}

\author{E.~Rzepniewski}
\affiliation{Department of Physics, University of Oxford,
The Clarendon Laboratory,
Parks Road, Oxford,~OX1~3PU, U.K.}

\author{R.D.~McDonald}
\affiliation{Department of Physics, University of Oxford,
The Clarendon Laboratory,
Parks Road, Oxford,~OX1~3PU, U.K.}

\author{J.~Schlueter}
\affiliation{Chemistry and
Materials Divisions, Argonne National Laboratory, 9700 South Cass
Avenue, Argonne, Illinois, U.S.A. }

\author{A.M.~Kini}
\affiliation{Chemistry and
Materials Divisions, Argonne National Laboratory, 9700 South Cass
Avenue, Argonne, Illinois, U.S.A. }

\begin{abstract}
Cyclotron resonance, along with its second and third harmonics, has
been observed in the quasi-two-dimensional organic superconductor
$\beta^{\prime\prime}$-(BEDT-TTF)$_2$SF$_5$CH$_2$CF$_2$SO$_3$. The
harmonic content is richly angle-dependent, and is interpreted in
terms of the interlayer warping of the Fermi surface giving rise to the
cyclotron resonance. In a departure from Kohn's theorem, but in
agreement with recent theoretical predictions, the effective mass
deduced from the cyclotron resonance measurements is greater than that
determined from magnetic quantum oscillations.

\end{abstract}


\maketitle

Many correlated-electron systems which are of fundamental or
technological interest have very anisotropic electronic bandstructure.
Examples include the ``high-$T_{\rm c}$'' cuprates~\cite{cuprates},
layered phases of the manganites~\cite{ramirez} and
ruthenates~\cite{ruthenate}, semiconductor superlattices~\cite{sl} and
crystalline organic metals~\cite{strong,review}.  Such systems are
often characterised by a tight-binding Hamiltonian in which the ratio
of the interlayer transfer integral $t_{\perp}$ to the average
intralayer transfer integral $t_{\parallel}$ is much less than 1.  The
resulting highly anisotropic dispersion relationships have been shown to
give rise to a range of interesting d.c.\ and a.c.\ angle-dependent
magnetoresistance oscillation (AMRO) effects~\cite{review}.

In this paper we describe an experiment demonstrating a new effect of
this kind, the variation with magnetic field orientation of the
harmonic content of cyclotron resonance (CR) measured in the
high-frequency interlayer conductivity. The effect arises because the
interlayer velocity component of a quasiparticle that executes a cyclotron
orbit around a warped cylindrical Fermi surface (FS) section acquires
higher harmonic content as the steady magnetic field causing the orbit
is tilted away from the cylinder axis. It may be considered a
high-frequency analogue of the Yamaji effect, in which the d.c.\
magnetoresistance oscillates as a function of magnetic field
orientation.

In a detailed study of the d.c.\ AMRO of the quasi-two-dimensional
(Q2D) organic molecular metal
$\beta^{\prime\prime}$-(BEDT-TTF)$_2$SF$_5$CH$_2$CF$_2$SO$_3$~\cite{geiser},
we use the Yamaji oscillations to parameterize the shape and
orientation of the Q2D FS of this material. Then, for the first 
time in any material 
the detailed angle-dependence of the harmonic content of CR is examined,
and we verify the form of the Fermi surface,
demonstrating that this effect could form the basis of a new technique
for studying the Fermi surface topology of low-dimensional
materials. The CR measurements also exhibit a mass renormalisation
effect recently predicted to occur in narrow-bandwidth metallic
systems~\cite{kanki}.

Experiments were carried out on single crystals ($\sim 0.5 \times 0.5
\times 0.1~{\rm mm}^3$) prepared by
electrocrystallisation~\cite{geiser}. Samples were initially
orientated using infrared reflectivity \cite{infrared} to $\pm
4^{\circ}$.  For d.c.\ measurements, electrical contacts were made
to the upper and lower faces (parallel to the {\bf ab} 
(Q2D) planes)
of selected crystals by attaching $25~\mu$m Au wires using graphite
paint (contact resistances $\ltsim 10~\Omega$).  The resistance was
measured by driving a low-frequency a.c.\ current ($5~\mu$A,
$17-200$~Hz) in the  ${\bf c}^*$ (interlayer) direction; the voltage was
measured in the ${\bf c}^*$ direction using a lock-in amplifier. In
this configuration, the resistance is proportional to the interplane
resistivity component $\rho_{zz}$~\cite{review}.  Samples were mounted
in a cryostat providing temperatures $T$ between 0.50~K and 4.2~K, and
allowing rotation about two mutually perpendicular axes. The sample's
angular coordinates are the angle $\theta$ between the magnetic field
{\bf B} and ${\bf c}^*$ and the angle $\phi$ defining the plane of
rotation~\cite{review,steve&m-s}.

\begin{figure}[htbp]
   \centering
   \includegraphics[height=13.8cm]{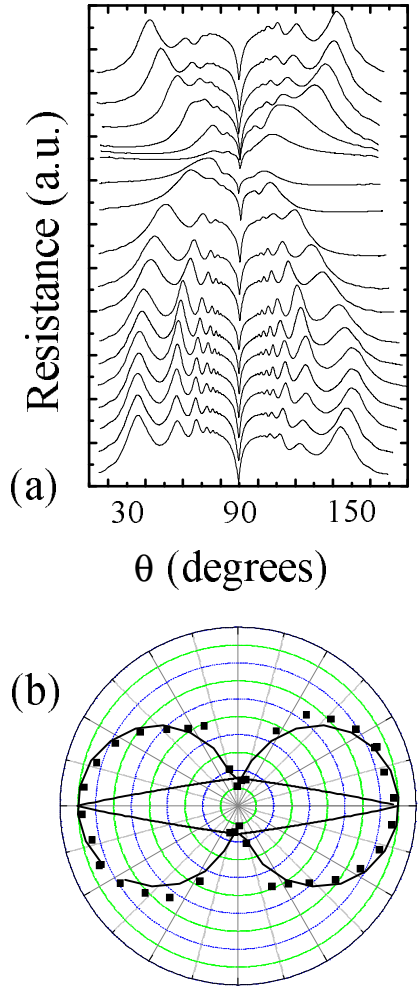}
    \caption{ \label{amrosw}
(a) AMRO data for
$\beta''$-(BEDT-TTF)$_2$SF$_5$CH$_2$CF$_2$SO$_3$ at 10~T and
1.5~K for $\phi$-angles $7\pm 1^{\circ}$ (top trace)
$17\pm 1^{\circ}$, $27 \pm 1^{\circ}$,
.....$177\pm 1^{\circ}$ (bottom trace-
adjacent traces spaced by
$10\pm 1^{\circ}$).
$\phi=0$ corresponds to
rotation in the {\bf a* c*} plane of the crystal
to within the accuracy of the infrared orientation.
(b)~The $\phi$ dependence of $k_{||}$
from the $\tan \theta$
periodicity of the AMRO in Fig.~1(a)
(points); the ``figure of eight'' solid
curve is a fit. The resulting fitted FS
pocket (elongated diamond shape; $j=1.1$) is shown
within. The long axis of the pocket makes an angle
of $68 \pm 4^{\circ}$ with the ${\bf b}^{\ast}$ axis.
}
\end{figure}

Fig.~\ref{amrosw}(a) shows the $\theta,\phi$ dependence of the
magnetoresistance for $B=10$~T and $T=1.5$~K.  The data show AMROs
with sharp maxima (the Yamaji effect), suggesting that they are due to
a Q2D FS section, rather than open sheets~\cite{review,steve&m-s}.
(For all $\phi$, a sharp dip is visible at $\theta=90^o$ as the 
superconducting upper critical field becomes large when
{\bf B} is parallel to the Q2D layers, causing a reduction in
$\rho_{zz}$.  See Ref.~\cite{janeloff} for a discussion.)  Such AMRO
maxima occur at angles $\theta_n$ given by $c^{\prime} k_\parallel
\tan(\theta_n)= \pi(n \pm \frac{1}{4})+f(\phi)$, where $c^{\prime}$ is
the effective interplane spacing, $n$ is an integer, $k_\parallel$ is
the maximum Fermi wave-vector in the plane of rotation and $f(\phi)$
is a function of the plane of rotation of the
field~\cite{review,steve&m-s}.  The $+$ and $-$ signs correspond to
$\theta_n >90^{\circ}$ and $\theta_n <90^{\circ}$ respectively.  The
$\phi$ dependence of the AMRO data is most accurately fitted by a
non-elliptical FS cross section~\cite{steve&m-s},
$(\frac{k_x}{k_a})^j+(\frac{k_y}{k_b})^j=1$.  A free-parameter fit
yielded $j=1.1 \pm 0.1$ and a ratio $k_a/k_b=9.0\pm 0.2$, leading to
the ``diamond-shaped'' FS shown in Fig.~\ref{amrosw}(b); the FS long
axis makes an angle of $68 \pm 4^{\circ}$ with the ${\bf b}^{\ast}$
axis of the crystal.  The fitted FS cross-section has an area
corresponding to a de Haas-van Alphen (dHvA) frequency of $F=196 \pm 3$~T, in excellent
agreement with Shubnikov-de Haas (SdH) and dHvA studies which yield $F=200 \pm
1$~T~\cite{wosdhva}. The disagreement between the area deduced from
AMRO and the dHvA experiments was significantly worse if the FS
cross-section was constrained to be elliptical ($j=2$)~\cite{wosdhva}.
Swept-field magnetotransport measurements were also carried out at
fixed $\theta, \phi$ over a range of $T$. A $\theta=0$ effective mass
$m^\ast$ of $1.96\pm 0.05~m_{\rm e}$ was obtained by fitting the
$T$-dependence of small-amplitude $F=200$~T SdH oscillations to the 2D
Lifshitz-Kosevich theory~\cite{review}, in good agreement with
previous studies~\cite{wosdhva}.

The CR measurements were made using resonant cavity
techniques~\cite{FTR}.  A
$\beta^{\prime\prime}$-(BEDT-TTF)$_2$SF$_5$CH$_2$CF$_2$SO$_3$ sample
was placed inside a rectangular cavity resonating in the TE$_{102}$
mode at 71.2~GHz~\cite{FTR}. Samples were aligned in the cavity such
that the oscillating {\bf H} field was parallel to the Q2D planes;
induced currents flow in the interplane (low-conductivity) direction,
leading to a skin depth greater than the sample
dimensions~\cite{FTR,newhill} and dissipation dominated by the interplane
conductivity~\cite{FTR,newhill,hillch}. The cavity can be rotated with
respect to {\bf B}, thus changing $\theta$ without changing the
electromagnetic environment of the sample in the cavity. The sample
can also be rotated inside the cavity to change $\phi$. Quasi-static 
fields were provided by
a superconductive magnet in Oxford and by a 33~T Bitter magnet at
NHMFL, Tallahassee.

\begin{figure}[htbp]
   \centering
   \includegraphics[height=10cm]{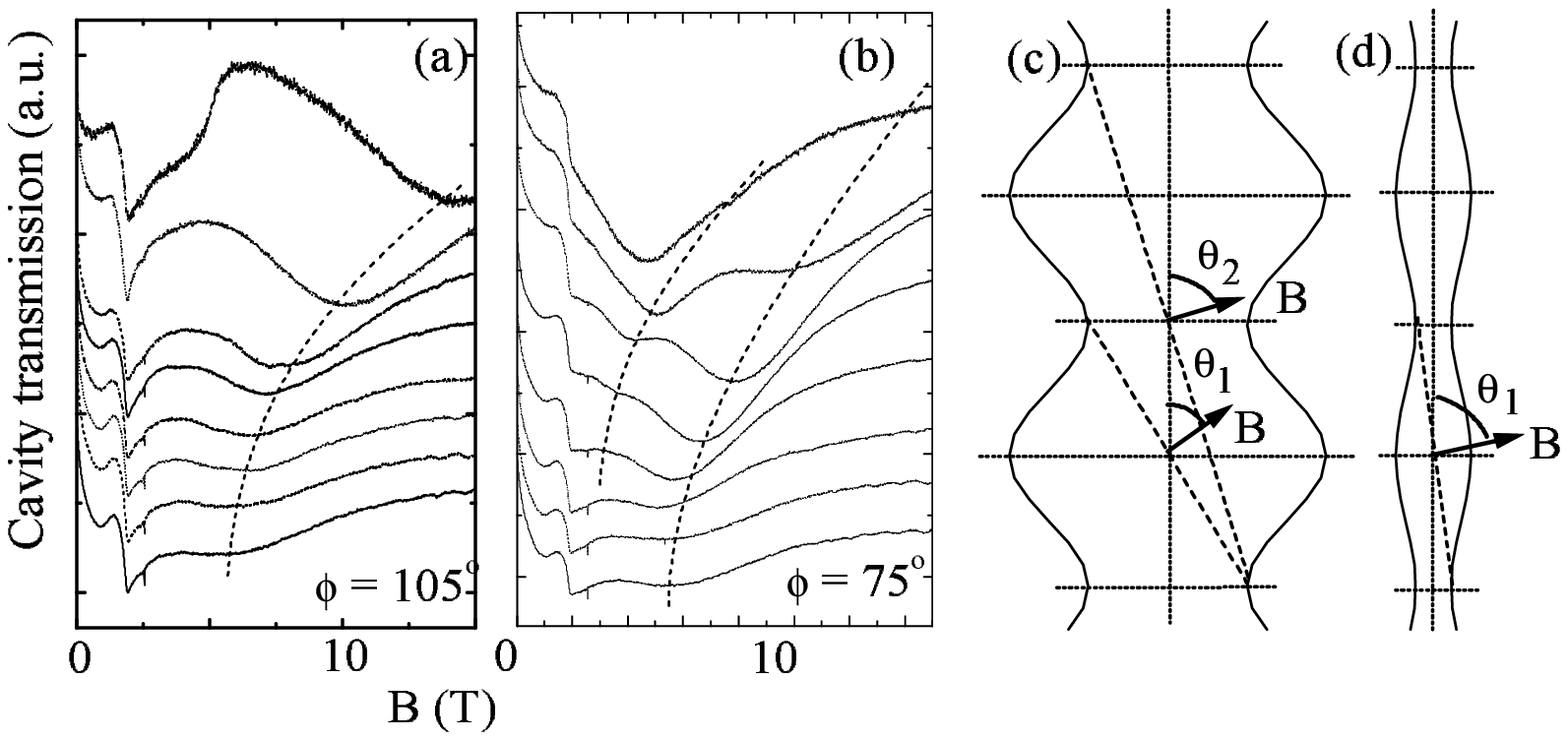}
\caption{  \label{mvnasw} (a) Transmission of the resonant
cavity loaded with a single crystal of
$\beta^{\prime\prime}$-(BEDT-TTF)$_2$SF$_5$CH$_2$CF$_2$SO$_3$
versus magnetic field for $\theta=0^{\circ}$ (lowest trace) to
$\theta=70^{\circ}$ (uppermost trace) and $\phi
=105^{\circ}$ ($T=1.5$~K).
(b) Equivalent data for $\phi = 75^{\circ}$.
(c) and (d)
represent two different cross-sections of a
warped cylindrical FS;
for clarity the warping has been very greatly
exaggerated. Cyclotron orbits
about the FS are shown schematically as
dotted lines for two inclinations of
the magnetic field
{\bf B} to the cylinder axis, $\theta_1$ and $\theta_2$.}
\end{figure}

Field sweeps were carried out at various $\theta$ for $\phi =
105^{\circ}, 75^\circ, 45^\circ$ and $15^\circ$. Fig.~\ref{mvnasw}
shows field sweeps between 0 and 15~T at a fixed frequency of
71.2~GHz, for $\phi = 105^\circ$ and $45^\circ$. Only $\theta$ angles
between $0$ and $70^\circ$ are shown, as the resonance positions are
symmetrical about $\theta=0$. The data have been normalised and offset
for clarity. The feature at 2~T on all sweeps is a background of the
apparatus~\cite{FTR}. At intermediate fields, broad resonances can be
seen. For $\phi=105^\circ$, one resonance moves to higher fields with
increasing angle. The position of the resonance in magnetic field for
different $\theta$ is fitted to $\frac{\omega}{B}=A\cos(\theta -
\theta_0)$~\cite{review}, giving values $A/\omega = 0.175 \pm~0.005$
T$^{-1}$ and $\theta_0=2^\circ \pm 2^\circ$. This is consistent with
the behaviour expected of a CR, where the resonance should be centred
on $\theta_0 = 0^\circ$~\cite{review}.

At $\phi=45^\circ$ (Fig.~\ref{mvnasw}(b)) a second resonance also appears
for higher $\theta$ at approximately half the field of the main
CR. This resonance is also seen for $\phi=75^{\circ}$ and
$\phi=15^{\circ}$, and also follows a $\cos\theta$ dependence of $1/B$ with
fitting parameters $A/\omega = 0.36~\pm~0.01$ T$^{-1}$ and $\theta_0$
= -1$^\circ$ $\pm$ 1$^\circ$. The value of $A/\omega$ strongly
suggests that this resonance is the second harmonic of the main CR.
At some azimuthal angles an additional feature becomes visible at
$\theta \sim 60^\circ$, behaving like a third harmonic of the CR.

\begin{figure}[tbp]
    \centering
     \includegraphics[height=13cm]{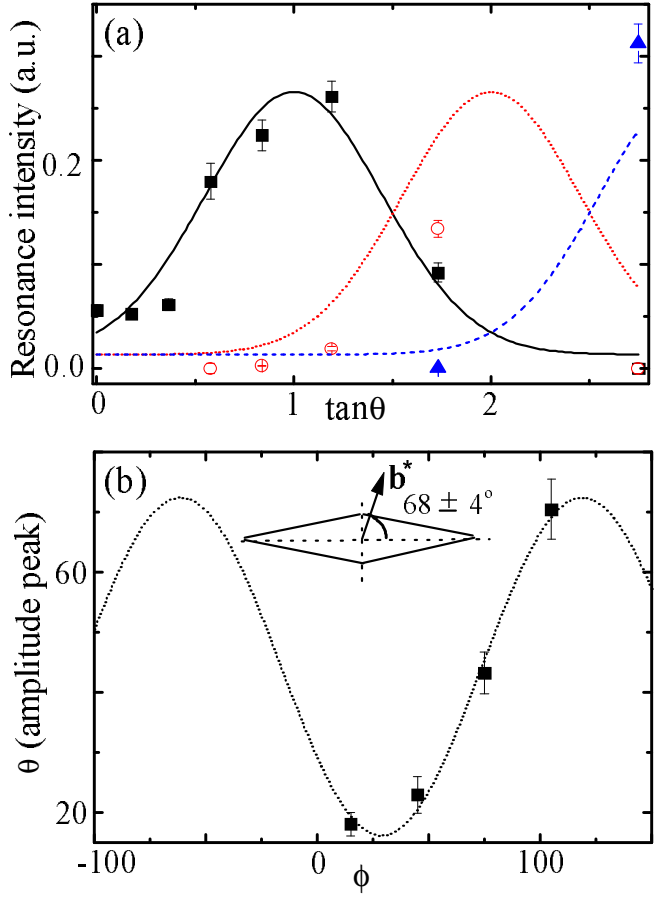}
    \caption{\label{mvnaFS} (a) Intensity
of CRs versus $\tan \theta$ for
$\phi=75^{\circ}$. Solid curve:
fit to the intensity
of the fundamental CR,
giving angle $\theta_1$
(data: square points).
Dotted curve: predicted intensity
of the second harmonic, centred on $\theta_2$
(data: hollow circles).
Dashed curve: predicted intensity of
the third harmonic, centred on $\theta_3$
(data: triangles).
(b) $\theta_1$ versus $\phi$;
points: experimental values;
the curve is a fit to Equation~\ref{phi}.
Inset: the orientation of the FS
cross-section at $68 \pm 4^{\circ}$
to the crystal's ${\bf b}^{\ast}$ axis derived from the fit.}
\end{figure}

Ref.~\cite{blundell} predicts the presence of harmonics in the real
space velocity of charge carriers in cyclotron orbits about
non-elliptical FS cross-sections, leading to harmonics of the CR.
It does not, however, predict even harmonics for a symmetrical FS such as that in
Fig.~\ref{amrosw}(b).  By contrast, Hill~\cite{hillch} considers a
cylindrical FS which is warped in the interplane direction, in which
the warping vector is {\it not} directed along the cylinder axis.
Harmonics of the CR are predicted, but the model is unable to
reproduce the behaviour reported here.  Instead we propose a mechanism
which predicts CR harmonics and their angular behaviour without the
need for a shifted warping vector. Fig.~\ref{mvnasw}(c) shows the
cross section of a warped Q2D FS section (warping greatly
exaggerated). For a general orientation of {\bf B}, quasiparticles
follow orbits about the FS such that the $z$-component of their real
space velocity, $v_z$ oscillates (the velocity is always perpendicular
to the FS~\cite{review}).  In our measurement geometry, the
dissipation caused by the sample is dominated by interplane currents,
i.e. by the behaviour of $v_z$~\cite{FTR,newhill}.  If $\theta$ is
small, the orbits remain within the length of one Brillouin zone (BZ)
in the $k_z$-direction and $v_z$ oscillates at the cyclotron frequency
$\omega_{\rm c}$. As $\theta$ increases, the orbits extend over
several BZs in the $k_z$ direction and $v_z$ acquires oscillatory
components at harmonics of $\omega_{\rm c}$. (Note that the precise
harmonic content depends on the average $k_z$ of the orbit; the
conductivity depends on an average over all possible
orbits~\cite{FTR,newhill}.) The largest amplitude of the $\omega_{\rm
c}$ resonance will occur at angle $\theta_1$ in Fig.~\ref{mvnasw}(c),
as the change in $v_z$ during an orbit will be greatest for all
possible orbits. The second harmonic will likewise be strongest at
$\theta_2$, and at $\theta_p$ for the $p$th harmonic.  Simple geometry
leads to
\begin{eqnarray}
\tan\theta_1 \approx \pi/ck_{\rm F};
~~~~~~\tan\theta_p \approx p\pi/ck_{\rm F}
\label{theta},
\end{eqnarray}
where $2 \pi/c$ is the height of the BZ in the interlayer direction
and $2k_{\rm F}$ is the average width of the FS cross-section in the
plane of rotation of {\bf B}.  Fig.\ref{mvnaFS}(a) shows the intensity
of the resonances at $\phi=75^{\circ}$ as a function of $\theta$.  The
$\theta$ dependence of the resonance intensity has been fitted to a
suitable functional form~\cite{edwards} centred on $\theta_1$.  The
angular behaviour of the second and third harmonics has then been
predicted using the same function, but centred on $\theta_2$ and
$\theta_3$ respectively.

The angle $\theta_1$ depends on the anisotropy of the FS and its
orientation. Figs.~\ref{mvnasw}(c) and (d) represent rotating the
field through the major and minor axes of the FS cross-section. In
Fig.~\ref{mvnasw}(d), $\theta_1$ is much higher than for (c).
$\theta_1$ should vary with rotation as
\begin{equation}
\theta_1=C+D\cos2(\phi+\phi_0), \label{phi}
\end{equation}
where $\phi_0$ is related to the orientation of the FS with respect to
the crystal axes and $C$ and $D$ are constants corresponding to the
major and minor axes of the FS cross-section. Fig.~\ref{mvnaFS}(b)
shows the values of $\theta_1$ for the $\phi$-orientations studied
fitted to Eqn.~\ref{phi}. From the maximum and minimum values of
$\theta_1$ the major and minor axes of the FS cross-section are found
to be in the ratio $10.5\pm 1.8$:1, in good agreement with the $9\pm
0.2$:1 found from the AMRO data.  The fitted value of $\phi_0$ is used
to orientate the FS with respect to the crystal axes, as shown in Fig.
\ref{mvnaFS}(b, inset). The orientation of the major axis (at $68 \pm
4^{\circ}$ to the ${\bf b}^{\ast}$ axis) agrees with orientations
deduced from the current and previous AMRO
data~\cite{steve&m-s,wosdhva} (also $68 \pm 4^{\circ}$).

A cyclotron mass $m^\ast_{\rm CR}=2.3~\pm~0.1~m_{\rm e}$ was deduced
from the angular frequency $\omega$ and field $B$ of the fundamental
CR at $\theta=0$ using $\omega = {eB}/{m^{\ast}_{\rm
CR}}$~\cite{review}.  CR gives a measure of the quasiparticle mass
different from that given by thermodynamic experiments such as the
dHvA effect~\cite{review,QBB}. In the simplest
picture~\cite{review,QBB}, the mass measured in CR represents the bare
band mass renormalised by electron-phonon interactions, whereas the
mass $m^{\ast}$ measured using quantum oscillatory techniques is
additionally renormalised by electron-electron interactions. In spite
of a large number of experiments on BEDT-TTF salts (see
Ref.~\cite{review} for details), only two definitive reports of CR,
both in $\alpha$-(BEDT-TTF)$_2$NH$_4$Hg(SCN)$_4$, have been
made~\cite{hillch,anton}. In these studies, a CR mass of
$m^{\ast}_{\rm CR}=1.9~m_{\rm e}$ was measured~\cite{hillch,anton};
this may be compared with $m^{\ast}=2.5~m_{\rm
e}$~\cite{marianoprl}. The relative sizes of the two masses
($m^\ast_{\rm CR} < m^\ast$) is therefore in agreement with the above
predictions.  By contrast, in the current experiments on
$\beta^{\prime\prime}$-(BEDT-TTF)$_2$SF$_5$CH$_2$CF$_2$SO$_3$,
$m^\ast=1.96 \pm 0.05m_{\rm e}$ and $m^\ast_{\rm CR}=2.3 \pm 0.1m_{\rm
e}$; {\it i.e.} $m^\ast_{\rm CR} > m^\ast$. Thus it appears that the
predictions of an enhancement of the effective mass derived from
quantum oscillations over that in a CR-like experiment do not
necessarily hold and that Kohn's theorem~\cite{kohn} is violated.
Hubbard-model calculations have recently been carried out~\cite{kanki}
which contradict the simple theories and appear to support the current
experimental data. In these calculations it was found that the
relationship between the two masses depends strongly on the
characteristics of the material measured (e.g.\ band filling) and in
some cases $m^\ast$ can be exceeded by $m^\ast_{\rm CR}$.

In summary, we have performed experiments on
$\beta^{\prime\prime}$-(BEDT-TTF)$_2$SF$_5$CH$_2$CF$_2$SO$_3$, showing
that the Q2D section of the Fermi surface (FS) is highly elongated and
non-elliptical.  Harmonics of cyclotron resonance (CR) have been
observed and are explained by a mechanism which confirms the
elongation and orientation of the FS cross-section.  The method is
consistent with a FS which is extended in the interlayer direction,
giving strong support for coherent interlayer charge-transfer in this
material. We have also found that the effective mass from CR
experiments is greater than that measured by magnetic quantum
oscillations, in agreement with recent theoretical work and in
violation of Kohn's theorem.

This work is supported by EPSRC (UK).  NHMFL Tallahassee is supported
by the US Department of Energy (DoE), NSF and the State of Florida. Work at
Argonne is sponsored by the DoE, Office of Basic Energy Sciences,
Division of Materials Science under contract number W-31-109-ENG-38.

\clearpage


\clearpage


\end{document}